\documentstyle[aps,prl,floats,psfig,twocolumn]{revtex}
\draft
\begin{document}
\title{Bose condensation in a model microcavity}

\author{P. R. Eastham and P. B. Littlewood} \address{Theory of
Condensed Matter, Cavendish Laboratory, Cambridge, CB3 0HE.  United
Kingdom.}

\date{\today} \maketitle \begin{abstract} We study the equilibrium
properties of a system of dipole-active excitons coupled to a single
photon mode at fixed total excitation. Treating the presence or
absence of a trapped exciton as a two-level system produces a model
that is exactly soluble. It gives a simple description of the physics
of polariton condensation in optical cavities beyond the low-density
bosonic regime.
\end{abstract} \pacs{71.35.Lk, 71.36.+c, 71.10.Li, 73.20.Dx}

\narrowtext The coupled exciton-photon modes of a microcavity were
first observed by Weisbuch et al.\cite{cavpol}. Since at low densities
excitons are bosons, at low densities coupled exciton-photon modes are
also bosons. These bosonic excitations are known as cavity polaritons.

Considerable effort has been devoted to seeking experimental evidence
for the bosonic nature of cavity polaritons. This evidence has been
sought in the non-equilibrium behavior of
microcavities\cite{senbloch,polstimdang,pau,paurecant,kira,tassone1,yura}. In
this paper we will consider the simpler equilibrium problem, in
particular the possibility of a Bose condensate of cavity polaritons.

Polaritons are not conserved particles, so there is ultimately no
equilibrium condensate. We may, however, treat polaritons as conserved
particles if their lifetime is much longer than the time required to
achieve thermal equilibrium at a fixed polariton number. We will study
such a quasi-equilibrium limit in a model microcavity.

While the theory of weakly-interacting bosons is well understood, it
is far from obvious that this theory is appropriate to the cavity
polariton condensate. The concept of a polariton\cite{hopfpol} is only
valid in linear response; it is not valid for a substantial occupation
of the excitons. Finite exciton densities introduce saturation of the
electronic states, so the excitons cannot be treated as bosons. In
general, finite densities of excitons also lead to exciton-exciton
interactions, which can produce dephasing and ionization of the
excitons\cite{exsatpol2}. By considering a situation in which neither
dephasing nor ionization are relevant, we will show how to generalize
the concept of a polariton to include saturation of the exciton
states. Saturation alone does not preclude Bose condensation.

We assume that the relevant electronic excitations in the microcavity
are localized, physically separated, single excitons. Thus we can
neglect the Coulomb interaction between excitons localized on
different sites. We further assume that, because of the tiny effective
mass of a cavity photon($\sim 10^{-5} m_{e}$ for a $1000$ \AA\
cavity), there is only a single relevant photon mode in the cavity.

These assumptions lead us to consider the well-known Dicke
model\cite{dickemodel}. This consists of a single mode of the photon
field, dipole coupled to a set of $N$ localized two-level
oscillators. Each two-level oscillator represents one exciton state,
localized on site $n$ with an energy $E_{g}(n)$. These exciton states
are composed of conduction and valence electrons with fermionic
annihilation operators $b_{n}$ and $a_{n}$ respectively; these
fermions are subject to the local constraints
$b^{\dagger}_{n}b_{n}+a^{\dagger}_{n}a_{n}=1$. For brevity, we
suppress the site index $n$ on the fermion operators. Making the
rotating wave approximation, we consider the Hamiltonian
\begin{eqnarray}
\label{ham}
H &=& \sum \frac{E_{g}(n)}{2} \left( b^{\dagger}b-a^{\dagger}a \right)
+ \omega_{c} \psi^{\dagger}\psi \\ & & + \frac{g}{\sqrt{N}}\sum \left(
b^{\dagger} a \psi + \psi^{\dagger} a^{\dagger} b \right). \nonumber
\end{eqnarray}
$\psi$ is the bosonic annihilation operator for the cavity mode and
the summations are over the site index $n$.

The operator $\sigma_{z}=\frac{1}{2}\sum (b^{\dagger}b - a^{\dagger}
a)$ measures the electronic excitation; such excitations are created
by the operator $\frac{1}{\sqrt{N}}\sum b^{\dagger}a$. They are
approximately bosonic provided we remain near to the bare ground
state, so that $\langle\sigma_{z}\rangle\approx-N/2$. In this limit
equation (\ref{ham}) becomes two coupled harmonic oscillators and we
recover the usual bosonic polaritons.

Away from $\langle\sigma_{z}\rangle=-N/2$ saturation of the
electronic states becomes relevant and the excitations are no longer
bosonic. However, we can still define a polariton: it is the quantum
of excitation of the non-linear coupled system. With this definition
the polariton number operator is $L=\psi^{\dagger} \psi + \sigma_{z}$,
which is a conserved quantity for the Hamiltonian (\ref{ham}).

The thermal equilibrium of the Dicke model was originally solved by
Hepp and Lieb\cite{hepplieb1}. Here we are interested in the
quasi-equilibrium problem posed by\ (\ref{ham}) at constant total
excitation $L$, a generalization obtained by adding a chemical
potential to constrain $L$. Thus at $T=0$ the relevant free energy is
$\langle H-\mu_{\mathrm ex}L\rangle$, where $\mu_{\mathrm ex}$ is the
chemical potential for excitations.

To keep the number of parameters to a minimum we restrict ourselves to
the uniform case $E_{g}(n)=E_{g}$. There are then only three
parameters in the problem: the temperature $(k_{B}\beta)^{-1}$, the
dimensionless detuning $\Delta=(\omega_{c}-E_{g})/g$, and the
excitation density $\rho_{\mathrm ex}=\langle L \rangle /N$.
Generalizing our solution to include a distribution of exciton
energies is straightforward.

Following Kiry'anov and Yarunin's work\cite{kiryan1} on the
unconstrained problem, we solve the model using the large-$N$
expansion of a coherent-state path-integral. This expansion is
possible because all the exciton states couple to a single mode of the
photon field. More detailed derivations, as well as a variational
approach based on the work of Comte and
Nozi\`{e}res\cite{nozieres:ex1} on the exciton condensate, will be
presented in a future publication.

\begin{figure}[t]
\centerline{\psfig{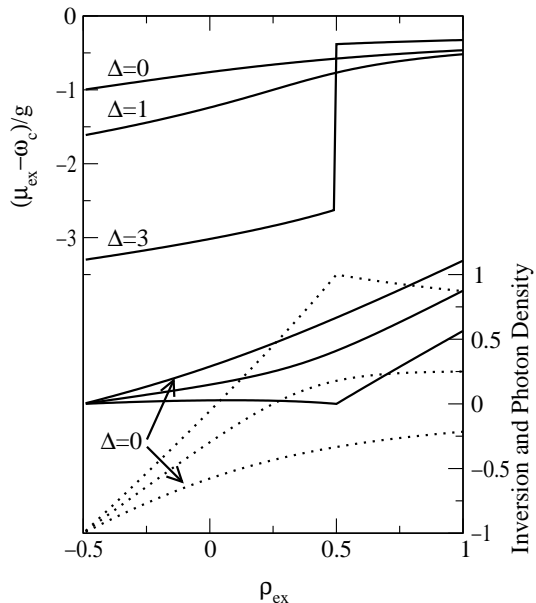}}
\caption{Behavior of the zero-temperature condensed solution as a
function of excitation density. Upper section (left axis): chemical
potential, for detunings $\Delta=0, 1$ and $3$. Lower section (right
axis): photon density(solid lines) and inversion density
$2\langle\sigma_{z}\rangle/N$(dotted lines), for the same three
detunings. The inversion increases with increasing $\Delta$, while
the photon density decreases.}
\label{fig1}
\end{figure}

By integrating out the fermions and rescaling the photon field, the
grand partition function for (\ref{ham}) can be represented as an
imaginary-time path-integral over the photon field of the form $Q
\propto \int{\cal D}\psi \exp (-N S_{\rm eff})$. In the thermodynamic
limit $N\to\infty$ the partition function is dominated by those
paths $\psi_{0}(\tau)$ which minimize the action $S_{\rm eff}$. The
Euler-Lagrange equation for $S_{\mathrm eff}$ is a self-consistency
condition which relates $\psi_{0}(\tau)$ to the equilibrium
polarization of a two-level system in the self-consistent field
$\psi_{0}(\tau)$. It takes the form
\begin{equation}
(\partial_{\tau}+\tilde{\omega}_{c})\psi_{0}(\tau)+
g\langle\bar{a}(\tau)b(\tau)\rangle=0,
\label{eq1}
\end{equation} with $\tilde{\omega_{c}}=\omega_{c}-\mu_{\mathrm ex}$. 

The order parameter of the polariton condensate is the polarization of
either the photon field or the exciton field. Although there are two
physically distinct order parameters, we see from equation\
(\ref{eq1}) that they are coupled by the dipole interaction. It is the
dipole interaction which favors the formation of a condensate.

We assume that the extremal trajectories are independent of $\tau$,
$\psi_{0}(\tau)=\psi_{0}$. We can then use the familiar
eigenstates\cite{gelesin} of a two-level system in a static external
field to calculate the polarization term on the right of equation\
(\ref{eq1}). A constrained thermal population of these renormalized
eigenstates leads to a BCS-like equation for $\psi_{0}$,
\begin{equation}
\tilde{\omega}_{c}\psi_{0}=\frac{g^2\psi_{0}}{2E}\tanh(\beta E).
\label{exeq}
\end{equation} Here $E$ is the renormalized fermion energy
$E=\sqrt{\tilde{\varepsilon}^2+g^2|\psi_{0}|^2}$, with
$\tilde{\varepsilon}=(E_{g}-\mu_{\mathrm ex})/2$.

The chemical potential $\mu_{\mathrm ex}$ is related to the partition
function $Q$ in the usual manner. For the condensed solutions the
leading term in the expansion of $Q$ around the extremal trajectories
gives
\begin{equation}
\label{denseq}
\rho_{\mathrm
ex}=|\psi_{0}|^2-\frac{1}{g^2}\tilde{\varepsilon}\tilde{\omega}_{c},
\end{equation} while for the normal solution we have $\rho_{\mathrm
ex}=-[\tanh(\beta\tilde{\varepsilon})]/2$.

At zero temperature (\ref{exeq}) and (\ref{denseq}) always have a
condensed solution $\psi_{0}\neq0$. This solution is illustrated in
Fig.\ \ref{fig1}. The upper part of Fig.\ \ref{fig1} shows the
chemical potential as a function of excitation density for detunings
$\Delta=0, 1$ and $3$. In the low density limit, $\rho_{\mathrm
ex}=-0.5$, we are describing a condensate of conventional bosonic
polaritons, so the chemical potential is given by the usual polariton
energy $\mu_{\mathrm
ex}=\frac{1}{2}\left((\omega_{c}+E_{g})-g\sqrt{\Delta^{2}+4}\right)$.
At high densities the electronic states are saturated and further
excitation must be added as photons. Thus the chemical potential
approaches the energy of the cavity mode. Between these two limits we
find a discontinuity at $\rho_{\mathrm ex}=0.5$ if $\Delta>2$. The
lower part of Fig. \ref{fig1} illustrates the composition of the
condensate, again for detunings $\Delta=0, 1$ and $3$. Increasing
detuning increases the electronic fraction of the condensate.  In the
high excitation limit the inversion approaches zero. This gives the
maximum polarization of the electronic states and hence minimizes the
dipole energy.

We study the stability of the solutions to (\ref{exeq}) by considering
the quadratic term $S_{2}$ in the functional Taylor series expansion
of $S_{\mathrm{eff}}$ around the extremal trajectory $\psi_{0}$. The
kernel of $S_{2}$, ${\mathcal{G}}^{-1}$, is the inverse matrix Green's
function for fluctuations of the photon field. We find
${\mathcal{G}}^{-1}$ by taking two functional derivatives of
$S_{\mathrm eff}$ and evaluating the result on the extremal trajectory
$\psi_{0}$. This amounts to solving the Dyson-Beliaev
equations\cite{popovbook} for the photon Green's functions. The
self-energies are provided by the polarizability of a two-level system
in the self-consistent field $\psi_{0}$.

The resulting expression for ${\mathcal G}^{-1}$ in the condensed
phase is somewhat lengthy, so we do not reproduce it here. We find
that the eigenvalues of ${\mathcal G}^{-1}$ in the condensed phase
are, apart from the Goldstone mode, strictly positive provided
$\tilde{\omega_{c}}>0$. This condition is automatically satisfied by
the condensed solutions to (\ref{exeq}).

To determine the phase diagram we assume a continuous transition
between the normal and condensed states. The transition temperature
$(k_{B}\beta_{c})^{-1}$ is determined by requiring (\ref{exeq}) and
(\ref{denseq}) to have a repeated root $\psi_{0}=0$. This gives two
transition temperatures
\begin{equation}
\label{phasebound}
\beta_{c}g=\frac{4\tanh^{-1}(2\rho_{\mathrm ex})}
{\Delta\pm\sqrt{\Delta^2-8\rho_{\mathrm ex}}}. 
\end{equation} We also obtain (\ref{phasebound})
as the temperature corresponding to the onset of a low frequency
instability of the normal state. Thus our assumption of a continuous
transition is correct.

\begin{figure}[t]
\centerline{\psfig{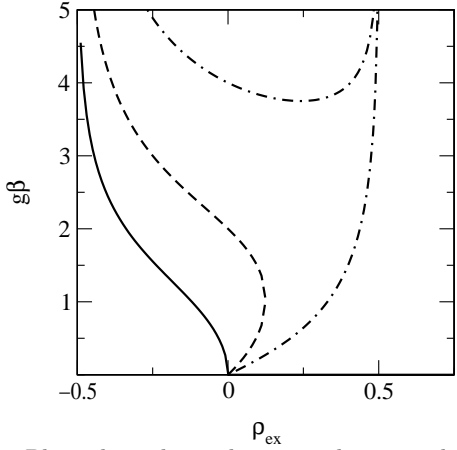}}
\caption{Phase boundaries between the normal and condensed states for
$\Delta=0$ (solid line), $\Delta=1$ (dashed line), and $\Delta=2$
(dot-dashed lines). The regions which include the lower left corner of
the figure are normal.}
\label{fig2}
\end{figure}

The phase boundary (\ref{phasebound}) is illustrated in
Fig. \ref{fig2} for detunings $\Delta=0, 1$ and $2$. For $\Delta\le0$
the phase diagram is straightforward. The transition temperature
increases monotonically with density, reaching infinity at
$\rho_{\mathrm ex}=0$. For $\Delta>0$ there is a region in which the
condensate exists on both the high and low temperature sides of the
normal state. When $\Delta\ge2$ the condensate exists in two
disconnected regions of the phase diagram.

This unusual phase diagram can be understood by considering the
excitations of the normal state. The excitation energies follow in the
usual manner from the locations of the poles of the Green's
function. We write the action for fluctuations of the photon field
about the normal state as
$S_{2}=\beta\sum_{\omega_{n}}\delta\bar\psi(\omega_{n}){\mathcal
G}^{-1}_{N}(\omega_{n})\delta\psi(\omega_{n})$, where $\omega_{n}$ is
a bosonic Matsubara frequency.  The normal-state Green's function
${\mathcal G}_{N}$ takes the form
\begin{equation}
{\mathcal{G}_{N}}(\omega_{n})=\frac{C_{+}}{i\omega_{n}+E_{+}}
+\frac{C_{-}}{i\omega_{n}+E_{-}},
\label{thgreen}
\end{equation} with
$E_{\pm}=[(\omega_{c}+E_{g})\pm g\sqrt{\Delta^2-8\rho_{\mathrm
ex}}]/2-\mu_{\mathrm ex}$ and
$C_{\pm}=\pm(2\tilde{\varepsilon}-E_{\pm})/(E_{-}-E_{+})$. The
excitation energies, measured from the chemical potential
$\mu_{\mathrm ex}$, are $E_{\pm}$.

The normal-state excitations are polaritons in the sense of
Hopfield\cite{hopfpol}: coupled modes involving the linear response of
the electronic system around its equilibrium state. The gap in the
spectrum is increased over the bare detuning $\Delta$ owing to the
dipole coupling between the excitons and the cavity mode.  The
presence of excitation in the ground state, either driven by finite
temperatures or by finite $\mu_{\mathrm ex}$, causes the two polariton
branches to attract. This attraction can be understood in terms of an
angular momentum representation\cite{dickemodel} for the collective
states of the electronic system. The excitation of the electronic
states $\sigma_{z}$ forms the z-component of an angular momentum and
their polarization forms a raising operator $\sigma_{+}$. Thus the
polarizability of the electronic states is a maximum at
$\langle\sigma_{z}\rangle=-N/2$.

\begin{figure}[t]
\centerline{\psfig{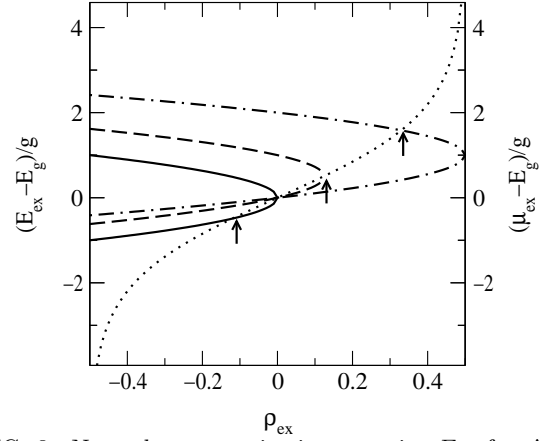}}
\caption{Normal state excitation energies $E_{\mathrm{ex}}$ for
$\Delta=0$ (solid curve), $\Delta=1$ (dashed curve), and $\Delta=2$
(dot-dashed curve), relative to the exciton energy $E_{g}$. The dotted
curve is the normal state chemical potential for $\beta g=1$ on the
same scale. The arrows mark the crossings which correspond to the
onset of condensation for these three detunings at $\beta g=1$.}
\label{fig3}
\end{figure}

Figure\ \ref{fig3} illustrates the excitation energies $E_{\mathrm
ex}=E_{\pm}+\mu_{\mathrm ex}$ obtained from equation (\ref{thgreen}),
for detunings $\Delta=0, 1$ and $2$. On the same axis we plot the
normal state chemical potential, given by the expression immediately
below equation (\ref{denseq}), for $\beta g=1$. Figure\ \ref{fig3}
should be compared with the $\beta g=1$ line of the phase diagram in
Fig. \ref{fig2}.

When $\Delta=0$ and $\rho_{\mathrm ex}=-0.5$ the system is in the
normal state. Increasing $\rho_{\mathrm ex}$ populates the electronic
excitations, increasing the chemical potential and decreasing the
polariton splitting. Eventually the chemical potential crosses the
lower polariton branch from below and the system condenses. In
contrast, for $\Delta=1$ and $2$ the chemical potential crosses the
lower polariton branch at $\rho_{\mathrm ex}=0$ without the condensate
appearing. It is not until the chemical potential crosses the upper
polariton branch that the transition occurs. This can be understood by
considering the signs of the quasiparticle weights $C_{\pm}$. A
positive quasiparticle weight corresponds to absorption of an external
field(particle-like excitations), whereas a negative quasiparticle
weight corresponds to gain(hole-like excitations). For $\rho_{\mathrm
ex}>0$, the lower polariton branch has a negative weight: it has
become hole-like, and must be below the chemical potential for
stability. Note that for large $\Delta$ there is a particle-hole
symmetry, so the region $-0.5<\rho_{ex}<0.5$ of the phase diagram is
symmetric about $\rho_{\mathrm ex}=0$.

In the condensed phase we find excitations with energies relative to
the chemical potential
$E_{\pm}=\pm\sqrt{(\tilde{\omega_{c}}+2\tilde{\epsilon})^{2}+
g^2|\psi_{0}|^2}$, along with a Goldstone mode. The polariton
condensate will produce incoherent luminescence or absorption at the
energies $E_{\pm}+\mu_{\mathrm ex}$ and coherent emission at
$\mu_{\mathrm ex}$.

Since we expect polariton condensates to produce coherent light, a
major issue experimentally will be distinguishing polariton
condensates from semiconductor lasers. In semiconductor lasers
excitons are conventionally assumed to be ionized. We list some
aspects of the present work which should help to distinguish polariton
condensation from lasing: (1)From Fig.\ \ref{fig1}, we see that the
polariton condensate can exist without an inversion of the electronic
system. (2)Approaching the condensation transition by increasing the
excitation density we may observe a reduction of the normal-state
excitation gap owing to the fermionic structure of the
excitons. (3)The fermionic structure of the excitons shifts the
coherent emission from the condensate away from the bosonic polariton
energy. (4)The incoherent luminescence and absorption from the
condensate exhibits a gap induced by the coherent photon field. In a
conventional laser, this gap is destroyed by the very short relaxation
time of the electronic polarization. Following \cite{calleg},
processes which destroy the electronic polarization could be
incorporated into our solution. In superconductors, such processes
produce a regime of gapless superconductivity\cite{natoconf1,parks}
where the order parameter survives without a gap in the excitation
spectrum.

The applicability of our results to real microcavities is restricted
by our neglect of exciton states with a significant wavefunction
overlap. We have assumed that these states are at infinitely large
energies. In reality, these states exist above an energy $E_{m}$. Thus
our thermodynamic results are only valid when $E_{m}-\mu_{ex}$ is
large compared with $\beta^{-1}$ and $g$. By considering Fig.\
\ref{fig3}, we deduce that realizing the phase diagram of Fig.\
\ref{fig2} requires an energy gap $\Delta E=E_{m}-E_{0} \gg g$. This
could occur in highly disordered materials with Frenkel-like excitons,
such as organic semiconductors\cite{organics,organics2}. An energy gap
$\Delta E$ could exist in inorganic quantum wells if the excitons move
in a potential containing deep, well-separated traps, perhaps
associated with interface
islands\cite{gammon1,hess,saturosc}. However, in both these cases it
is likely that there will be several exciton states on each site,
rather than the single state we have assumed.

If $\Delta E$ is small compared with $g$ then the phase boundary
(\ref{phasebound}) will only be realized when $\Delta \ll 0$ and the
temperature is low. In the normal state under these conditions both
the exciton occupation and the effects of the dipole interaction are
negligible. The transition temperature (\ref{phasebound}) then
corresponds to non-interacting bosons with an unusual density of
states.

To summarize, we have presented a theory of polariton condensation in
the Dicke model. By studying the model at fixed excitation, we have
generalized the concept of a polariton condensate from the low-density
regime. We have found two states: a normal state of excitons and a
condensate of polaritons. The polariton condensate is a superposition
of a BCS-like state of the excitons and a coherent state of the photon
field, and is favored over the normal state by the dipole
coupling. We recover conventional polaritons both as the low-density
limit of the condensate and as the linear-response excitations of the
normal state.

This work was supported by funding from the Engineering and Physical
Sciences Research Council, UK.

\bibliographystyle{prsty}

\end{document}